\documentclass[%
 aip,
 amsmath,amssymb,
 reprint,%
]{revtex4-1}

\usepackage{graphicx}% Include figure files
\usepackage{dcolumn}% Align table columns on decimal point
\usepackage{bm}% bold math
\usepackage[utf8]{inputenc}
\usepackage[T1]{fontenc}
\usepackage{mathptmx}
\usepackage{etoolbox}
\usepackage{hyperref}
\hypersetup{
colorlinks=true,
linkcolor=blue,
anchorcolor=blue,
citecolor=blue}
\makeatletter
\def\@email#1#2{%
 \endgroup
 \patchcmd{\titleblock@produce}
  {\frontmatter@RRAPformat}
  {\frontmatter@RRAPformat{\produce@RRAP{*#1\href{mailto:#2}{#2}}}\frontmatter@RRAPformat}
  {}{}
}%
\makeatother
\begin{document}

\preprint{AIP/123-QED}

\title[]{Lopsided Rigid Dumbbell Rheology from Langevin Equation: \\
A Graduate Tutorial}
% Force line breaks with \\
\author{Nhan Phan-Thien}
\affiliation{State Key Laboratory of Fluid Power and Mechatronic Systems and Department of Engineering Mechanics, Zhejiang University, Hangzhou 310027, China%\\This line break forced with \textbackslash\textbackslash
}%
\affiliation{Department of Mechanical Engineering, National University of Singapore, Singapore 117575, Singapore}

\author{Dingyi Pan}
\affiliation{State Key Laboratory of Fluid Power and Mechatronic Systems and Department of Engineering Mechanics, Zhejiang University, Hangzhou 310027, China%\\This line break forced with \textbackslash\textbackslash
}%
\email{dpan@zju.edu.cn}

\author{Mona A. Kanso}%
\affiliation{Chemical Engineering Department, Massachusetts Institute of Technology, Cambridge, Massachusetts 02139, USA%\\This line break forced with \textbackslash\textbackslash
}%

\author{Alan Jeffrey Giacomin}
\affiliation{Mechanical Engineering Department, University of Nevada, Reno, Nevada 89557, USA}
\affiliation{State Key Laboratory for Turbulence and Complex Systems, Peking University, Beijing 100871, China} 

%\\This line break forced% with \\
%}%

\date{\today}% It is always \today, today,

\begin{abstract}
The modelling of symmetric rigid dumbbell particles suspended in a Newtonian fluid, as a model of a rigid-rod polymeric solution, has been accomplished exclusively through the diffusion equation, which has been detailed elegantly by Bird et al. [Chapter 14 of \textit{Dynamics of Polymeric Liquids}, Vol 2, Ed 2, (1987)]. In this tutorial, a straightforward approach for modelling a lopsided rigid dumbbell particle is presented by the Langevin analysis. The connector force between the dumbbell beads is obtained through the rigidity constraint of the center-to-center vector of the dumbbell using its Langevin equation. By directly averaging via the Langevin equation, the evolution of the center-to-center vector, and the configuration tensor are derived. The stress expressions for the dumbbell from the Langevin equation, and the diffusion equation for the orientation distribution function of the center-to-center vector of the dumbbell are also derived, and the final expressions agree with established results from other methods. 
\end{abstract}

\maketitle

\section{Lopsided Rigid Dumbbell} 
Kinetic macromolecular theory, following the elegant method of Refs.\citenum{Bird,BWE,BirdGiacomin,BirdArmstrong}, always proceeds in two steps.  First, we solve for the orientation distribution function, then we integrate in phase space (subject to normalization) to get rheological material functions. This Tutorial is for those teaching this elegant method. Within this Tutorial we arrive at the Giesekus expression for the stress, in a completely different way, starting with the Langevin equation. Those learning the dynamics of polymeric liquids will see the equivalence between kinetic macromolecular theory and the Langevin approach. Whereas the kinetic molecular theory approach requires the derivation of an orientation distribution function, the Langevin approach does not.  Moreover, students can see that the results of the Langevin approach apply exactly for any choice of the dumbbell statistics, including the orientation distribution.
In this Tutorial we rework the specific example of a lopsided rigid dumbbell.  Though our work on this is mainly driven by curiosity, its many practical applications have not escaped our attention. Lopsided dumbbell suspensions arise in the lipid encapsulation of mRNA into vaccines (FIGURE 3 of Ref.~\citenum{Brader}), meningococcal infections (Fig. 1.B of Ref.~\citenum{Manno}), or when certain viral proteins interact with membranes (see Figures 1. through 3. of Ref.~\citenum{Alimohamadi}), or when synthesized deliberately from plastics. \cite{Peng,KimWoong} 

Our work has focused on two beads and is silent on more general bead-rod structures with differently-sized beads. The general rigid bead-rod theory, for identically sized beads, is now well-developed (EXAMPLE 16.7-1 of Ref. \citenum{Bird} (or EXAMPLE 13.6-1 of Ref. \citenum{BirdArmstrong}). \cite{Kanso,KansoBook,Hassager1,Hassager2,Kansothesis}  The canonical example of shish-kebabs with differently sized beads is also untouched. We leave these intriguing advanced problems as exercises for the student. 

In this Tutorial, we support the equations with specific citations to textbooks [\textit{Dynamics of Polymeric Liquids}, Vol 2, Ed 2, (1987)~\cite{Bird} and 
\textit{Understanding Viscoelasticity - An Introduction to Rheology}, Ed 3, (2017)~\cite{NPT}]. The Tutorial is meant to enrich the learning experiences of those following either approach.
%add specific citations above
\subsection{Langevin Equations}

\begin{figure}
    \centering
    \includegraphics[width=\linewidth,trim=11cm 3cm 11cm 3cm,clip]{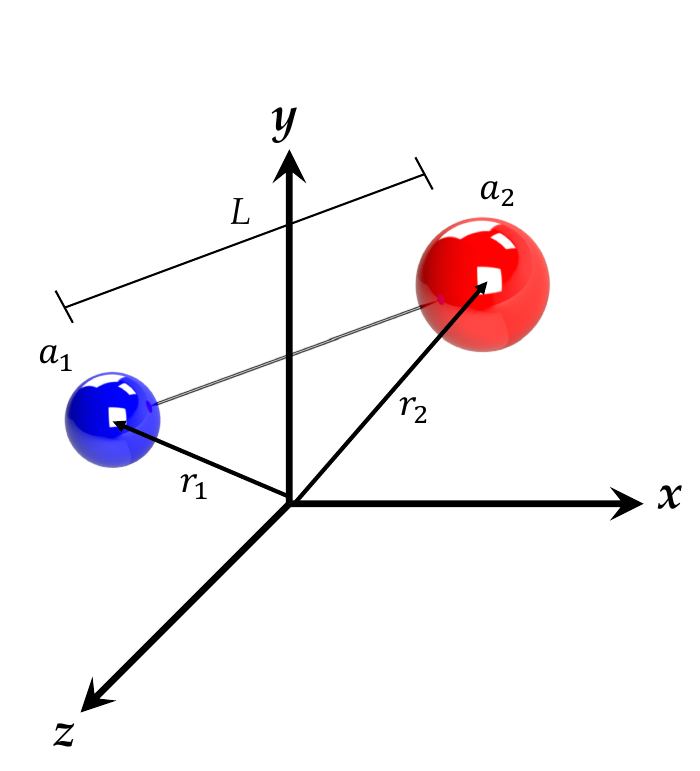}
    \caption{Lopsided rigid dumbbell model ($\nu=2$).}
    \label{fig:1}
\end{figure}

\begin{table*}[tb]
	\caption{Nomenclature and Symbols.}\label{tab:1}
	\begin{ruledtabular}
	\begin{tabular}{lcc}
	Name & Unit & Symbol\\
	\hline
Acceleration, $i$th bead & $L/t^2$ & $\ddot{\mathbf{r}}_i$ \\
Bead radius, $i$th bead  &  $L$ & $a_i$ \\
Bead Stokes friction coefficient & $M/t$ & $\zeta$ \\
Bead Stokes friction coefficient, $i$th bead &$M/t$ & $\zeta_i$ \\
Boltzmann temperature & $ML^2/t^2$ &	$kT$ \\
Diffusivity, rotational & $L^2/t$ & $D$ \\
Dirac delta function & $t^{-1}$  & $\delta$ \\
Dumbbell center of resistance & $L$ & $\mathbf{R}^{(c)}$ \\
Dumbbell center-to-center distance, rigid dumbbell & $L$ & $L$ \\
Dumbbell center-to-center vector & $L$ & $\mathbf{R}$ \\
Force of connector, between rigid dumbbell beads & $ML/t^2$ & $\mathbf{F}^{(c)}$ \\
Force on center of resistance, Brownian	& $ML/t^2$ & $\mathbf{F}^{(b,c)}\left(t\right)$ \\
Force on center-to-center vector, Brownian & $ML/t^2$ & $\mathbf{F}^{(b,\mathbf{R})}\left(t\right)$ \\
Force on $i$th bead, random Brownian & $ML/t^2$ & $\mathbf{F}_{i}^{(b)}\left(t\right)$ \\
Identity tensor & $1$ & $\mathbf{I}$ \\
Lopsidedness, dumbbell & $1$ & $\nu$ \\
Number density, dumbbells & $L^{-3}$ & $n$ \\
Position vector, $i$th bead & $L$ & $\mathbf{r}_i$ \\
Probability density distribution function & $L^{-3}$ & $\phi$ \\
Stress tensor, dumbbell contribution to	& $M/Lt^2$ & $\mathbf{S}^{\left(p\right)}$ \\
Time & $t$ & $t$ \\
Time, interval of past & $t$ & $s$ \\
Unit vector pointing from bead $1$ to bead $2$ &	$1$ & $\mathbf{p}$ \\
Vector from the center of resistance to bead $i$ & $L$ & $\mathbf{R}^{(i)}$ \\
Velocity gradient gradient, at center of resistance & $L^{-1}t^{-1}$ & $\nabla\nabla\mathbf{u}^{(c)}$ \\
Velocity gradient, at center of resistance & $t^{-1}$ & $\nabla\mathbf{u}^{(c)}$ \\
Velocity gradient, at the center of resistance, transpose of & $t^{-1}$ & $\mathbf{L}$ \\
Velocity, fluid, at $i$th bead & $L/t$ & $\mathbf{u}_i$ \\
Velocity, $i$th bead & $L/t$ & $\dot{\mathbf{r}}_i$ \\
Viscosity, solvent & $M/Lt$ & $\eta_s$ \\
	\end{tabular}
	\end{ruledtabular}
\begin{flushleft}
Legend: $M\equiv$mass; $L\equiv$length; and $t\equiv$ time.
\end{flushleft}
\end{table*}

We consider a lopsided rigid dumbbell suspended in a Newtonian solvent (Fig.~\ref{fig:1}), where all the interactions between the solvent and the dumbbell beads are located at the bead centers, at position vectors $\mathbf{r}_{1}$ and $\mathbf{r}_{2}$. Table~\ref{tab:1} gives nomenclature, symbols and dimensions for our variables, both dimensional and dimensionless. The center-to-center distance between the beads, $L$, is fixed. The beads are thus connected by a \textit{phantom} rigid rod. By \textit{phantom}, we mean massless, dimensionless, and that the rod does not interfere in any way with the Stokes flows of the beads to which it connects. Each bead $i=1,2$ is associated with a radius $a_{i}$, a Stokes frictional coefficient
\begin{equation}
\zeta_{i}=6\pi\eta_{s}a_{i},
\label{Eq1}%
\end{equation}
with $\eta_{s}$ the solvent viscosity, and a negligible mass $m_{i}$. Let $i=2$ designate the second of the two beads, and let the dumbbell lopsidedness be the ratio
\begin{equation}
\nu=\frac{a_{2}}{a_{1}}.
\end{equation}
 and thus, $\zeta_2=\nu \zeta_1$.  So, if $a_2>a_1$, then $\nu>1$ also.  Results for the symmetric dumbbell must therefore be recovered when $\nu=1$.
 
To maintain a constant distance between the beads, a connector force $\mathbf{F}^{(c)}$ between them arises; $\mathbf{F}^{(c)}$ is thus a constraint on the dynamical system to keep the dumbbell rigid. The beads undergo their Newtonian 2nd law of motions acted on by a fluid viscous resistance force, $\zeta_{i}\left(\mathbf{u}_{i}-\dot{\mathbf{r}}_{i}\right)$, with $\mathbf{u}_{i}=\mathbf{u}\left(\mathbf{r}_{i}\right)$ the fluid velocity evaluated at the bead $i$, a constraint connector force between the two beads, $\pm\mathbf{F}^{(c)}$, and a random force $\mathbf{F}_{i}^{(b)}\left(t\right)$ modelled by Brownian motion ($i=1,2$) representing the random impulses from the solvent molecules on the beads: 
\begin{equation}%
\begin{array}
[c]{c}%
m_{1}\ddot{\mathbf{r}}_{1}=\zeta_{1}\left(\mathbf{u}_{1}-\dot{\mathbf{r}}_{1}\right) +\mathbf{F}^{(c)}+\mathbf{F}_{1}^{(b)}\left(t\right),\\
m_{2}\ddot{\mathbf{r}}_{2}=\zeta_{2}\left(\mathbf{u}_{2}-\dot{\mathbf{r}}_{2}\right) -\mathbf{F}^{(c)}+\mathbf{F}_{2}^{(b)}\left(t\right).
\end{array}
\label{1}%
\end{equation}

In the dynamical Langevin system Eq.~(\ref{1}),~\cite{Langevin} the Brownian forces $\mathbf{F}_{i}^{(b)}$ are modelled by white noise. We define the center-to-center vector and its unit vector $\mathbf{p}$, pointing from bead 1 to bead 2, 
\begin{equation}
\mathbf{R}=L\mathbf{p}=\mathbf{r}_{2}-\mathbf{r}_{1},%
\end{equation}
where $L$ is the center-to-center distance, a constant distance between beads 1 and 2. The Brownian forces have zero means, uncorrelated to each other, and their strength (or autocorrelation function) is given by a fluctuation-dissipation theorem relating their strength to the mobility of the beads (see Landau and Lifshitz~\cite{Landau} for many types of fluctuation-dissipation theorems, and also in Subsection 7.3 of Phan-Thien and Mai-Duy~\cite{NPT}):%
\begin{equation}
\begin{aligned}
\langle\mathbf{F}_{j}^{(b)}\left(t\right)\rangle &= \boldsymbol{0},\\
\langle\mathbf{F}_{1}^{(b)}\left(t+s\right)\mathbf{F}_{2}^{(b)}\left(t\right)\rangle &= \boldsymbol{0},\\
\langle\mathbf{F}_{1}^{(b)}\left(t+s\right)\mathbf{F}_{1}^{(b)}\left(t\right)\rangle &= 2kT\zeta_{1}\delta\left(s\right)\mathbf{I},\\
\langle\mathbf{F}_{2}^{(b)}\left(t+s\right)\mathbf{F}_{2}^{(b)}\left(t\right)\rangle &= 2kT\zeta_{2}\delta\left(s\right)\mathbf{I}.
\end{aligned}
\label{2}%
\end{equation}
where $\delta\left(s\right)$ is the delta Dirac function, $\mathbf{I}$ is the identity tensor, $kT$ is the Boltzmann temperature representing the kinetic energy of the beads, and the chevrons denote an ensemble average operations.

Let us now define the center of resistance $\mathbf{R}^{(c)}$ by
\begin{equation}
(\zeta_{1}+\zeta_{2})\mathbf{R}^{(c)}=\zeta_{1}\mathbf{r}_{1}+\zeta_{2}\mathbf{r}_{2}.
\end{equation}
By \textit{center of resistance} , we mean the point on the dumbbell where a single applied force results in pure translation. Since $\zeta_{2}=\nu\zeta_{1}$, the center of resistance is also defined by $\left(1+\nu\right)  \mathbf{R}^{(c)}=\mathbf{r}_{1}+\nu\mathbf{r}_{2}$.  We also denote by $\mathbf{R}^{(i)}$ the vector from the center of resistance to bead $i$:
\begin{equation}
\begin{gathered}
\mathbf{r}_{1}=\mathbf{R}^{(c)}-\mathbf{R}^{(1)},\;\mathbf{r}_{2}=\mathbf{R}^{(c)}+\mathbf{R}^{(2)},\\
\mathbf{R}=\mathbf{R}^{(1)}+\mathbf{R}^{(2)},%
\end{gathered}
\end{equation}
Then, by virtue of the definition of the center of resistance,
\begin{equation}
\zeta_{2}\mathbf{R}^{(2)}-\zeta_{1}\mathbf{R}^{(1)}=\zeta_{1}\mathbf{r}_{1}+\zeta_{2}\mathbf{r}_{2}-(\zeta_{1}+\zeta_{2})\mathbf{R}^{(c)}=\mathbf{0}.%
\label{6}%
\end{equation}

On the time scales of interest the inertia of the beads can be ignored, as it is customarily done. The equations of motion for the center of resistance, and for the center-to-center vector become,
\begin{equation}
\begin{gathered}
\left(\zeta_{1}+\zeta_{2}\right)\dot{\mathbf{R}}^{(c)}=\zeta_{1}\mathbf{u}_{1}+\zeta_{2}\mathbf{u}_{2}+\mathbf{F}^{(b,c)}\left(t\right),\\
\zeta_{2}\dot{\mathbf{R}}=\zeta_{2}\left(\mathbf{u}_{2}-\mathbf{u}_{1}\right)  -\frac{\zeta_{1}+\zeta_{2}}{\zeta_{1}}\mathbf{F}^{(c)}+\mathbf{F}^{(b,\mathbf{R})}\left(  t\right),
\end{gathered}
\label{3}%
\end{equation}
where $\mathbf{F}^{(b,c)}\left(t\right)$ is the Brownian force acting on the center of resistance, and $\mathbf{F}^{(b,\mathbf{R})}\left(t\right)$ is the Brownian force acting on the center-to-center vector; they are given by 
\begin{equation}
\begin{aligned}
\mathbf{F}^{(b,c)}\left(t\right) & =\mathbf{F}_{1}^{(b)}\left(t\right)+\mathbf{F}_{2}^{(b)}\left(t\right),\\
\mathbf{F}^{(b,\mathbf{R})}\left(t\right) & =\mathbf{F}_{2}^{(b)}\left(t\right)-\nu\mathbf{F}_{1}^{(b)}\left(t\right).
\end{aligned}\label{4}%
\end{equation}

Furthermore a Taylor's series expansion about the center of resistance in $\mathbf{u}$ yields%
\begin{equation}
\begin{aligned}
\mathbf{u}_{1} &  =\mathbf{u}^{(c)}-\mathbf{R}^{(1)}\cdot\nabla\mathbf{u}^{(c)}+\mathbf{R}^{(1)}\mathbf{R}^{(1)}\!:\!\nabla\nabla\mathbf{u}^{(c)}\!+\!HOT,\\
\mathbf{u}_{2} &  =\mathbf{u}^{(c)}+\mathbf{R}^{(2)}\cdot\nabla\mathbf{u}^{(c)}+\mathbf{R}^{(2)}\mathbf{R}^{(2)}\!:\!\nabla\nabla\mathbf{u}^{(c)}\!+\!HOT,
\end{aligned}
\end{equation}
where $HOT$ are terms of higher orders, and the superscript $(c)$ refers to an evaluation at the center of resistance. Consequently,
\begin{equation}
\begin{aligned}
\zeta_{1}\mathbf{u}_{1}+\zeta_{2}\mathbf{u}_{2}&=\left(  \zeta_{1}+\zeta_{2}\right)  \mathbf{u}^{(c)}+\zeta_{1}\mathbf{R}^{(1)}\mathbf{R}^{(1)}:\nabla\nabla\mathbf{u}^{(c)}\\
&+\zeta_{2}\mathbf{R}^{(2)}\mathbf{R}^{(2)}:\nabla\nabla\mathbf{u}^{(c)}+HOT,
\end{aligned}
\end{equation}
and
\begin{equation}
\begin{aligned}
\mathbf{u}_{2}-\mathbf{u}_{1}&=\mathbf{L}\cdot\mathbf{R+R}^{(2)}\mathbf{R}^{(2)}:\nabla\nabla\mathbf{u}^{(c)}\\
&-\mathbf{R}^{(1)}\mathbf{R}^{(1)}:\nabla\nabla\mathbf{u}^{(c)}+HOT,
\end{aligned}
\end{equation}
where $\mathbf{L}=(\nabla\mathbf{u}^{(c)})^T$ is the velocity gradient at the center of resistance. Thus
\begin{equation}
\begin{aligned}
\dot{\mathbf{R}}^{(c)} &  =\mathbf{u}^{(c)}+\left(  \zeta_{1}+\zeta_{2}\right)^{-1}\mathbf{F}^{(b,c)}\left(t\right),\\
\dot{\mathbf{R}} &  =\mathbf{L\cdot R}-\frac{\zeta_{1}+\zeta_{2}}{\zeta_{1}\zeta_{2}}\mathbf{F}^{(c)}+\zeta_{2}^{-1}\mathbf{F}^{(b,\mathbf{R})}\left(t\right),
\label{8}%
\end{aligned}
\end{equation}
where the statistics of the Brownian forces acting on the center of resistance and the center-to-center vector can be found \textcolor{black}{from} Eq.~(\ref{2}):%
\begin{equation}
\begin{aligned}
\langle\mathbf{F}^{(b,c)}\left(  t\right)  \rangle &  =\boldsymbol{0}, \\
\langle\mathbf{F}^{(b,c)}\left(  t+s\right)  \mathbf{F}^{(b,c)}\left(t\right)  \rangle &  =2kT\left(  \zeta_{1}+\zeta_{2}\right)  \delta\left(s\right)  \mathbf{I},\\
\langle\mathbf{F}^{(b,\mathbf{R})}\left(  t\right)  \rangle &  =\boldsymbol{0},\\
\langle\mathbf{F}^{(b,\mathbf{R})}\left(  t+s\right)  \mathbf{F}^{(b,\mathbf{R})}\left(t\right)  \rangle &  =2kT\zeta_{2}\left(  \frac{\zeta_{1}+\zeta_{2}}{\zeta_{1}}\right)  \delta\left(  s\right)  \mathbf{I}, \label{13}%
\end{aligned}
\end{equation}
Because of the rigidity constraint, the connector force on the right hand of Eq.~(\ref{8}) must satisfy $\mathbf{R\cdot \dot{R}}=0=\mathbf{p\cdot \dot{R}}$; thus%
\begin{equation}
\begin{aligned}
\mathbf{L}:\mathbf{Rp}-\frac{\zeta_{1}+\zeta_{2}}{\zeta_{1}\zeta_{2}}\mathbf{p}\cdot \mathbf{F}^{(c)}+\zeta_{2}^{-1}\mathbf{p}\cdot \mathbf{F}^{(b,\mathbf{R})}\left(  t\right)  =0,
\end{aligned}
\end{equation}
and since $\mathbf{F}^{(c)}$ can only have component parallel to $\mathbf{p}$, we can express the connector force as%
\begin{equation}
\mathbf{F}^{(c)}=\frac{\zeta_{1}\zeta_{2}}{\zeta_{1}+\zeta_{2}}\mathbf{L}:\mathbf{Rpp}+\frac{\zeta_{1}}{\zeta_{1}+\zeta_{2}}\mathbf{pp}\cdot\mathbf{F}^{(b,\mathbf{R})}\left(  t\right).\label{16}%
\end{equation}
Compared with the {\it symmetric} dumbbell case, where the deterministic part of the connector force is $\mathbf{F}^{(c)}=\zeta\mathbf{L}:\mathbf{Rpp}/2$, the effect of lopsidedness is seen to replace the effective coefficient of friction by
\begin{equation}
\frac{\zeta}{2}=\frac{\zeta_{1}\zeta_{2}}{\zeta_{1}+\zeta_{2}}\ \mathrm{or\ }\frac{2}{\zeta}=\frac{1}{\zeta_{1}}+\frac{1}{\zeta_{2}}.\label{18}%
\end{equation}
That is, the effective frictional coefficient is the harmonic mean of the beads' frictional coefficients, a result noted by Abdel-Khalik and Bird.~\cite{Khalik} Alternatively, since the frictional coefficient is proportional to the radius, this amounts to replacing both radii by an effective radius $a$ given by the harmonic mean of the radii (Eq.~(103) of  Ref.~\citenum{Poungthong}),
\begin{equation}
\frac{a}{2}=\frac{a_{1}a_{2}}{a_{1}+a_{2}}\ \mathrm{or\ }\frac{2}{a}=\frac{1}{a_{1}}+\frac{1}{a_{2}}.
\label{19}
\end{equation}

In addition, the rigidity constraint also demands that only Brownian forces acting perpendicular to the center-to-center vector participate in the equation of motion for $\mathbf{R}$, the components parallel to $\mathbf{R}$ cause a drift in the mean square of the center of resistance and are of no interest; the filter $\left(  \mathbf{I}-\mathbf{pp}\right)$ on the Brownian force $\mathbf{F}^{(b,\mathbf{R})}\left(  t\right)  $ effectively does that; $\left(  \mathbf{I}-\mathbf{pp}\right)  \mathbf{F}^{(b,\mathbf{R})}\left(  t\right)  $ can be justifiably called the rotational Brownian force, and the Langevin equation for the center-to-center vector is%
\begin{equation}
\dot{\mathbf{R}}=\mathbf{L}\cdot\mathbf{R}-\mathbf{L}:\mathbf{ppR}+\zeta_{2}^{-1}\left(\mathbf{I}-\mathbf{pp}\right)  \mathbf{F}^{(b,\mathbf{R})}\left(  t\right)  , \label{7}%
\end{equation}
 which we will use presently to generate the diffusion equation for the center-to-center vector, from which entities related to $\mathbf{R}$ can be averaged. 

\subsection{Fokker--Planck Equation}
The statistics of any random variable can only be considered fully known when its probability distribution is fully specified,\cite{Bird} and it is of pedagogical interest to derive this diffusion, or Fokker--Planck equation\cite{Fokker}$^-$\cite{Planck} for the \textcolor{black}{stochastic} process $\mathbf{R}\left(t\right)$ from its Langevin equation Eq.~(\ref{7}). For \textcolor{black}{any} stochastic process, \textcolor{black}{and particularly for $\mathbf{R}\left(t\right)$, Eq.~(\ref{7})}, Chandrasekhar~\cite{chandrasekhar} showed that its probability density distribution function $\phi\left(\mathbf{R},t\right) $ satisfies the following diffusion, or Fokker-Planck equation (also given in Subsection 7.3 of Phan-Thien and Mai-Duy~\cite{NPT})
\begin{equation}
\frac{\partial\phi}{\partial t}=\lim_{\Delta t\rightarrow0}\frac{\partial}{\partial\mathbf{R}}\left[  \frac{\left\langle \Delta\mathbf{R}\Delta\mathbf{R}\right\rangle }{2\Delta t}\frac{\partial\phi}{\partial\mathbf{R}}-\frac{\left\langle \Delta\mathbf{R}\right\rangle }{\Delta t}\phi\right].
\end{equation}
Now, from Eq.~(\ref{7}) and to order $\Delta t$,%
%\begin{align*}
\begin{equation}
\begin{aligned}
\Delta\mathbf{R} & =\left(\mathbf{L}\cdot\mathbf{R}-\mathbf{L}:\mathbf{ppR}\right)  \Delta t\\
 & +\zeta_{2}^{-1} \int_{t}^{t+\Delta t}\left( dt \mathbf{I}-\mathbf{pp}\right) 
\mathbf{F}^{(b,\mathbf{R})}\left(  t\right),
\end{aligned}
\end{equation}
%\end{align*}
and consequently%
\begin{equation}
\left\langle \Delta\mathbf{R}\right\rangle =\left(  \mathbf{L}\cdot\mathbf{R}-\mathbf{L}:\mathbf{ppR}\right)  \Delta t.
\end{equation}
Furthermore, keeping to $O\left(  \Delta t\right)  $ and using subscript notation for clarity,%
\begin{align}
\left\langle \Delta R_{i}\Delta R_{j}\right\rangle  &  =\zeta_{2}^{-2}\int%
_{t}^{t+\Delta t}\int_{t}^{t+\Delta t}dtd\tau\left(  \delta_{ik}-p_{i}%
p_{k}\right)  \left(  \delta_{jm}-p_{j}p_{m}\right)  \nonumber \\
& \hspace{26mm} \cdot\left\langle
F_{k}^{(b,\mathbf{R})}\left(  t\right)  F_{m}^{(b,\mathbf{R})}\left(  \tau\right)  \right\rangle
\nonumber\\
&  =2kT\frac{\zeta_{1}+\zeta_{2}}{\zeta_{1}\zeta_{2}}\left(  \delta_{ij}%
-p_{i}p_{j}\right).
\end{align}
Thus the Fokker--Planck equation \textcolor{black}{for the stochastic process $\mathbf{R}(t)$, Eq.~(\ref{7}),} becomes%
\begin{equation}
\begin{aligned}
\frac{\partial\phi}{\partial t} &  =\frac{\partial}{\partial\mathbf{R}%
}\cdot\left[  kT\frac{\zeta_{1}+\zeta_{2}}{\zeta_{1}\zeta_{2}}\left(
\mathbf{I}-\mathbf{pp}\right)  \frac{\partial\phi}{\partial\mathbf{R}}\right.\\
&  \left.  \phantom{\frac{\partial}{\partial\mathbf{R}} }-\left(
\mathbf{LR}-\mathbf{L}:\mathbf{ppR}\right)  \phi\right].
\end{aligned}\label{25}
\end{equation}
\textcolor{black}{Noting that $\mathbf{R} = L\mathbf{p}$}, the rotational diffusivity of the \textcolor{black}{stochastic} process $\mathbf{p}(t)$ is
\begin{equation}
D=2kT\frac{\zeta_{1}+\zeta_{2}}{\zeta_{1}\zeta_{2}L^{2}}=\frac{{4}kT}{\zeta L^{2}}.
\end{equation}
\textcolor{black}{The diffusion equation Eq.~(\ref{25})} is exactly as given by Bird et al.\cite{Bird} (with a correct interpretation of the operator $\partial/\partial\mathbf{R}$, or $L^{-1}\partial/\partial\mathbf{p}$ on a unit sphere), when the frictional factors are replaced by their harmonic means, as noted by Abdel-Khalik and Bird.\cite{Khalik}

\subsection{Average Motions and Configuration Tensor}

If the flow is homogeneous, i.e., $\mathbf{L}$ is constant, $\nabla\nabla\mathbf{u}^{(r)}$ and $HOT$ are identically zero, the dumbbell's center of resistance drifts just like a particle of fluid in the mean,
\begin{equation}
\langle\dot{\mathbf{R}}^{(c)}\rangle=\langle\mathbf{u}^{(c)}\rangle=\mathbf{L}\langle\mathbf{R}^{(c)}\rangle=\mathbf{u}\left(  \langle\mathbf{R}^{(c)}\rangle\right).
\end{equation}
A migration from the streamline of the center of resistance is \textcolor{black}{therefore} induced by a non-homogeneous flow field (or by \textcolor{black}{other} external forces on the beads, \textcolor{black}{not considered here}). This migration is proportional to $\nabla\nabla\mathbf{u}^{(c)}$ and the $HOT$.

Also, the average center-to-center vector evolves in time according to 
\begin{equation}
\langle\dot{\mathbf{R}}\rangle=\mathbf{L}\cdot\langle\mathbf{R}\rangle-\mathbf{L}:\langle\mathbf{ppR}\rangle. \label{10}%
\end{equation}
This equation does not involve a material time constant, unlike in the elastic dumbbell case -- the only time constant comes from the shear rate tensor. To solve for the average motion, higher-order statistics in $\mathbf{R}$ are required, a classical closure problem.

Multiplying Eq.~(\ref{7}) by $\mathbf{R}$, averaging the resulting expression, the evolution of the configuration tensor $\langle\mathbf{RR}\rangle$ is found,%
\begin{align}
\frac{d}{dt}\langle\mathbf{R}\mathbf{R}\rangle  &=\left\langle \mathbf{R}\dot{\mathbf{R}}+\mathbf{\dot{R}R}\right\rangle  \label{5} \\
&=\mathbf{L}\cdot\left\langle \mathbf{RR}\right\rangle +\left\langle \mathbf{RR}\right\rangle\cdot\mathbf{L}^{T}-2\mathbf{L}:\left\langle \mathbf{ppRR}\right\rangle \nonumber\\
&+\zeta_{2}^{-1}\left\langle \mathbf{R}\left(  \mathbf{I-pp}\right)\mathbf{F}^{(b,\mathbf{R})}\left(  t\right)  +\left(  \mathbf{I-pp}\right)\mathbf{F}^{(b,\mathbf{R})}\left(  t\right)  \mathbf{R}\right\rangle. \nonumber
\end{align}
The average of quantity like $\mathbf{R}\left(  \mathbf{I}-\mathbf{pp}\right)\mathbf{F}^{(b,\mathbf{R})}\left(  t\right)  $ can be found by (see, for example, Subsection 7.5 of Phan-Thien and Mai-Duy~\cite{NPT})%
\begin{equation}
\left\langle \mathbf{R}\left( \mathbf{I}-\mathbf{pp}\right)  \mathbf{F}^{(b,\mathbf{R})}  \right\rangle 
=\left\langle \left[\mathbf{\mathbf{R}}  \left(  \mathbf{I}-\mathbf{pp}\right)  \mathbf{F}^{(b,\mathbf{R})}\right]_{ \Delta t}\right\rangle ,\label{17}%
\end{equation}
where the subscript $\Delta t$ on an entity denotes an evaluation at time $\Delta t$. With $\mathbf{R}\left(\Delta t\right)$ provided by integrating Eq.~(\ref{7}) and making use of vastly different time scales between kinematic quantities and the Brownian forces, we obtain (using subscripts for clarity, and showing only non-trivial results),%
%remark: Dingyi - the $zeta^{-1}$ arises from integrating (19), which is then canceled out in forming the average as indicated in (15). Note that the integral of the Dirac delta function from 0 to Delta t returns 1/2 (only half of the full value of 1. end remark
\begin{align}
\left\langle R_{i}\left(  \Delta t\right)  F_{j}^{(b,\mathbf{R})}\left(  \Delta t\right)  \right\rangle &=\zeta_{2}^{-1}\left\langle \left(  \delta_{ik}-p_{i} p_{k}\right) \phantom{\int_{0}^{\Delta t}}\right. \nonumber\\
&\hspace{3mm}\left.\times\int_{0}^{\Delta t}\!\!F_{k}^{(b,\mathbf{R})}\left(  t\right)  F_{j}^{(b,\mathbf{R})}\left(  \Delta t\right)  dt\right \rangle \nonumber\\
&=kT\left(  1+\nu\right)  \left\langle \delta_{ij}-p_{i}p_{j}\right\rangle,\label{11}%
\end{align}%
\textcolor{black}{
%\begin{equation}
\begin{align}
\left\langle \left[R_{i}p_{j}p_{k}  F_{k}^{(b,\mathbf{R})}\right]_{\Delta t}  \right\rangle &=\zeta_{2}^{-1}\left\langle p_{i}p_{j}\left(  \delta_{km}-p_{k}p_{m}\right) \phantom{\int_{0}^{\Delta t}} \right.\nonumber\\
&\hspace{3mm}\left.\times\int_{0}^{\Delta t}\!\!F_{k}^{(b,\mathbf{R})}\left(  t\right)  F_{m}^{(b,\mathbf{R})}\left(  \Delta t\right)  dt \right\rangle\qquad\nonumber\\
&=2kT\left(  1+\nu\right)  \left\langle p_{i}p_{j}\right\rangle ,\label{12}%
\end{align}
%\end{equation}
}\\
using the statistics from Eq.~(\ref{13}) and the property of the Dirac delta function. \textcolor{black}{The results of Eqs.~(\ref{11}--\ref{12}) are used in Eq.~(\ref{5})}, and with $\mathbf{R}=L\mathbf{p}$, we get
\begin{equation}
\begin{aligned}
\frac{d}{dt}\langle\mathbf{RR}\rangle &-\mathbf{L}\cdot\left\langle \mathbf{RR}\right\rangle -\left\langle \mathbf{RR}\right\rangle\cdot\mathbf{L}^{T}+2\mathbf{L}:\left\langle \mathbf{ppRR}\right\rangle \\
&+\frac{6kT\left(  1+\nu\right)  }{\zeta_{2}L^{2}}\left\langle \mathbf{RR}\right\rangle =\frac{2kT\left(  1+\nu\right)  }{\zeta_{2}}\mathbf{I}.%
\end{aligned}
\end{equation}
Now, defining the time constant
\begin{equation}
\lambda=\frac{\zeta_{2}L^{2}}{6kT\left(  1+\nu\right)  }=\frac{\zeta_{1}\zeta_{2}L^{2}}{6kT\left(  \zeta_{1}+\zeta_{2}\right)  }=\frac{\zeta L^{2}}{12kT},
\end{equation}
with the equivalent friction factor $\zeta$ from Eq.~(\ref{18}),
we have
\begin{equation}
\begin{gathered}
\lambda\left\{  \frac{d}{dt}\langle\mathbf{pp}\rangle-\mathbf{L}\cdot\left\langle\mathbf{pp}\right\rangle -\left\langle \mathbf{pp}\right\rangle\cdot\mathbf{L}^{T}+2\mathbf{L}:\left\langle \mathbf{pppp}\right\rangle \right\}  \\
+\left\langle \mathbf{pp}\right\rangle =\frac{1}{3}\mathbf{I},%
\end{gathered}
\end{equation}
which is exactly the evolution given in Bird et al.~\cite{Bird} At equilibrium, $\left\langle \mathbf{pp}\right\rangle =\frac{1}{3}\mathbf{I},$ as it must. Denoting the upper convected derivative on a second-order tensor $\mathbf{A}$ by%
\begin{equation}
\frac{\delta}{\delta t}\mathbf{A}=\frac{d}{dt}\mathbf{A}-\mathbf{LA}%
-\mathbf{AL}^{T},
\end{equation}
then the evolution of the ensemble average configuration tensor $\left\langle \mathbf{pp}%
\right\rangle $ is%
\begin{equation}
\lambda\frac{\delta}{\delta t}\langle\mathbf{pp}\rangle+2\lambda
\mathbf{L}:\left\langle \mathbf{pppp}\right\rangle +\left\langle
\mathbf{pp}\right\rangle =\frac{1}{3}\mathbf{I.}%
\end{equation}

\section{Stress from Langevin equation}
The stress contributed by the dumbbells is related to the configuration tensor by (Eq.~D of Table 13.3-1 of Bird et al.~\cite{Bird})

\begin{equation}
\begin{aligned}
\mathbf{S}^{\left(  p\right)  } &  =-nkT\mathbf{I}-3nkT\lambda\frac{\delta}{\delta t}\left\langle \mathbf{pp}\right\rangle \\
&  =-nkT\mathbf{I}+
6nkT\lambda\mathbf{L}:\left\langle \mathbf{pppp}\right\rangle +3nkT\left\langle \mathbf{pp}\right\rangle \label{14}%
\end{aligned}
\end{equation}

The first form in Eq.~(\ref{14}) is known as the Giesekus expression (Eq. D of Table 13.3-1 of Ref. \citenum{Bird}),~\cite{Giesekus} and the second, the Kramers expression (Eq. D of Table 13.3-1 of Ref. \citenum{Bird}),~\cite{Kramers} for the dumbbell-contributed stress.

It is pedagogical to derive the stress expressions from Langevin equation of motion. The convenient starting point is the expression for the stress, from Bird et al.~\cite{Bird} (Eq.~13.3-14) specialized to this case:%
\begin{equation}
\mathbf{S}^{(p)}=n\left\langle \mathbf{RF}^{(c)}\right\rangle -\frac{1}{2}n\left\langle \mathbf{RF}_{2}^{(b)}-\mathbf{RF}_{1}^{(b)}\right\rangle.\label{15}%
\end{equation}
The first term on the right of Eq.~(\ref{15}) is called the contribution of intramolecular forces (due to the connector force), and the second term, the contribution of external forces (in this case they consist of Brownian forces) by Bird et al.~\cite{Bird} With $\mathbf{F}^{(c)}$ given by Eq.~(\ref{16}), we have
%\begin{equation}
%\mathbf{F}^{(c)}=\frac{\zeta_{1}\zeta_{2}}{\zeta_{1}+\zeta_{2}}\mathbf{L}:\mathbf{Rpp}+\frac{\zeta_{1}}{\zeta_{1}+\zeta_{2}}\mathbf{pp}\cdot\mathbf{F}^{(b,R)}\left(  t\right),
%\end{equation}
%

\begin{equation}
\begin{aligned}
\left\langle \mathbf{RF}^{(c)}\right\rangle & =\frac{\zeta L^{2}}{2}\left\langle\mathbf{L}:\mathbf{pppp}\right\rangle\\
& +\frac{\zeta_{1}R}{\zeta_{1}+\zeta_{2}}\left\langle \mathbf{ppp}\cdot\mathbf{F}^{(b,\mathbf{R})}\left(  t\right)  \right\rangle.
\end{aligned}
\end{equation}
Using the technique as detailed in Eqs.~(\ref{17}-\ref{12}), and $\zeta L^{2}/2=6kT\lambda$, we find that
\begin{equation}
\left\langle \mathbf{RF}^{(c)}\right\rangle =6kT\lambda\left\langle \mathbf{L}:\mathbf{pppp}\right\rangle +2kT\left\langle \mathbf{pp}\right\rangle ,\label{20}%
\end{equation}
and%
\begin{align}
\left\langle \mathbf{RF}_{2}^{(b)}\right\rangle  & =\zeta_{2}^{-1}\left\langle\int_{0}^{\Delta t}\left(  dt \mathbf{I}-\mathbf{pp}\right)  \mathbf{F}^{(b, \mathbf{R})}\left(  t\right)  \mathbf{F}_{2}^{(b)}\right\rangle \nonumber\\
& =\zeta_{2}^{-1}\left\langle \int_{0}^{\Delta t}dt \left( \mathbf{I}-\mathbf{pp}\right)  \left(  \mathbf{F}_{2}^{(b)}-\nu\mathbf{F}_{1}^{(b)}\right)  \mathbf{F}_{2}^{(b)}\right\rangle \nonumber\\
& =kT\left(  \mathbf{I}-\mathbf{pp}\right)  =-\left\langle \mathbf{RF}_{1}^{(b)}\right\rangle.
\label{19a}%
\end{align}
Putting results Eqs.~(\ref{20}-\ref{19a}) into Eq.~(\ref{15}), the dumbbell-contributed stress becomes (Eq.~A of Table 13.3-1 of Ref.~\citenum{Bird})%
\begin{align}
\mathbf{S}^{(p)}  & =6nkT\lambda\left\langle \mathbf{L}:\mathbf{pppp}\right\rangle +2kT\left\langle \mathbf{pp}\right\rangle -nkT\left(\mathbf{I}-\mathbf{pp}\right)  \nonumber\\
& =-nkT\mathbf{I}+3kT\left\langle \mathbf{pp}\right\rangle +6nkT\lambda \left\langle \mathbf{L}:\mathbf{pppp}\right\rangle , \label{35}
\end{align}
which is exactly the Kramers expression for the stress in Eq.~(\ref{14}). This, using the evolution of the configuration tensor and the definition of the upper-convected derivative, leads to the Giesekus expression Eq.~(\ref{14}) (Eq.~D of Table 13.3-1 of Ref.~\citenum{Bird}).

Rather than starting with Eq.~(\ref{15}), one can start with the volume average concept for the stress, as detailed in Subsection 7.4 of {Phan-Thien and Mai-Duy},~\cite{NPT} wherein it is set as an exercise for the student to derive the dumbbell-contributed stress as given exactly by Eq.~(\ref{35}) (Eq.~A of Table 13.3-1 of Ref.~\citenum{Bird}).  

%add specific citations to the underlined

\section{Conclusion}
A lopsided rigid dumbbell model is developed by Langevin analysis of Brownian motion. The connector force between the dumbbell beads is obtained as a constraint to the rigidity requirement, and it is found that a harmonic mean of the friction coefficients (Eq.~(\ref{18})), or the harmonic mean of the radii (Eq.~(\ref{19})), of the lopsided beads can be used to replace the effective coefficient of friction of the symmetric rigid dumbbell model -- all the previously derived results for the symmetric rigid dumbbell then apply with this replacement, a result uncovered by Abdel-Khalik and Bird using kinetic molecular theory  (Rows 2 of TABLE 13.4-1 of Ref.~\citenum{BirdHassager} or TABLE 16.4-1 of Ref.~\citenum{Bird}, Ref.~\citenum{Khalik}).

The Langevin equation for the center-to-center vector is then derived. By averaging the center-to-center vector, the evolution of the configuration tensor is found. The expressions for the dumbbell-contribution stress are also derived from the Langevin approach, and our final expressions Eqs.~(\ref{15}) and (\ref{35}) are exactly those derived from the distribution function, detailed in Bird et al.~\cite{Bird} The diffusion equation for the probability distribution in the center-to-center vector can be derived from its Langevin equation. %\st{The centre-to-center vector is often named unfortunately the end-to-end vector. This point is moot as the beads are just point masses - their sizes are a reflection of their Stokes resistances. The terminology may be a hangover from the chain model for polymer, with elasticity from the random walk of the many segments between the end points. As a reminder, we are not dealing with a physical dumbbell here - only point Stokes resistances separated by a fixed distance.}  
We consider this Tutorial a pedagogical contribution -- indeed the Langevin analysis may bring out the role of lopsidedness more clearly, especially in the case where deterministic external forces are acting on the beads~\cite{Kim} (not considered in this Tutorial, and for which Eq.~(\ref{15}) may need to be modified). Our Langevin approach will be preferred by those having no interest in the orientation distributions of the macromolecules.  It is less work. Our Langevin approach will also be preferred by students with engineering mechanics background. One can \textcolor{black}{definitely} have just as much fun with the Langevin approach for analyzing lopsided elastic dumbbells (see Ref.~\citenum{NhanKanso}).

\section*{ACKNOWLEDGMENT}
D. Pan thanks the National Natural Science Foundation of China (Grant No. 12222211) for support. This teaching and research was also undertaken, in part, thanks to support from the Discovery Grant program of the Natural Sciences and Engineering Research Council of Canada (NSERC) Tier 1 Canada Research Chair in Physics of Fluids (A.J. Giacomin).

\section*{AUTHOR DECLARATIONS}
\subsection*{Conflict of Interest}
The authors have no conflicts to disclose.

\section*{Data Availability Statement}

The data that support the findings of this study are available within the article.

\section*{References}

\bibliography{aipsamp}% Produces the bibliography via BibTeX.

\end{document}